\documentclass[aps,twocolumn,a4paper,pra,nofootinbib]{revtex4}
\usepackage{amssymb,amsmath}
\usepackage{psfig}
\usepackage{latexsym}
\usepackage{amsfonts}
\usepackage{graphics}
\usepackage{graphicx}

\begin{document}

\title{Electromagnetic vacuum fluctuations,\\
Casimir and Van der Waals forces}
\author{Cyriaque Genet}
\altaffiliation[Present address: ]{Huygens Laboratory, Universiteit Leiden, 
P.O. Box 9504, 2300 RA Leiden The Netherlands} 
\author{Francesco Intravaia}
\author{Astrid Lambrecht}
\author{Serge Reynaud}
\email{reynaud@spectro.jussieu.fr}
\homepage{www.spectro.jussieu.fr/Vacuum}
\affiliation{Laboratoire Kastler Brossel, case 74, 
Campus Jussieu, 75252 Paris, France}
\thanks{Laboratoire du CNRS, de l'ENS 
et de l'Universit{\'e} Pierre et Marie Curie} 
\date{Final version, 18 September 2003}

\begin{abstract}
Electromagnetic vacuum fluctuations have observable consequences, like the
Casimir force between mirrors in vacuum. This force is now measured with
good accuracy and agreement with theory when the effect of imperfect
reflection of mirrors is properly taken into account. We discuss the simple
case of bulk metallic mirrors described by a plasma model and show that
simple scaling laws are obtained at the limits of long and short distances.
The crossover between the short and long-distance laws is quite similar
to the crossover between Van der Waals and Casimir-Polder forces for
two atoms in vacuum.
The result obtained at short distances can be understood as the London
interaction between plasmon excitations at the surface of each bulk mirror.
\end{abstract}

\maketitle

\section{Introduction}

An important prediction of quantum theory is the existence of irreducible
fluctuations of electromagnetic fields even in vacuum, that is in the
thermodynamical equilibrium state with a zero temperature. These
fluctuations have a number of observable consequences in microscopic
physics for example in atomic physics the Van der Waals force
between atoms in vacuum.

Vacuum fluctuations also have observable mechanical effects in macroscopic
physics and the archetype of these effects is the Casimir force between two
mirrors at rest in vacuum. This force was predicted by H. Casimir in 1948
\cite{Casimir48} and soon observed in different experiments which confirmed
its existence \cite{Sparnaay89,Milonni94,Mostepanenko97,LamoreauxResource99}.
Recent experiments have reached a good precision, in the \% range, which
makes possible an accurate comparison between theoretical predictions and
experimental observations \cite{Bordag01,Lambrecht02}.

Such a comparison is important for at least two reasons. On one hand, the
Casimir force is a mechanical consequence of vacuum fluctuations which raise
a serious challenge at the frontier of quantum theory with the physics of
gravitation; it is therefore important to test this prediction with the
greatest care and accuracy. On the other hand, several experiments are
searching for hypothetical new forces predicted by the models of unification of
fundamental forces and the main target of these experiments is now the
distance range between the nanometer and millimeter; in this distance range,
the Casimir force is the dominant interaction between neutral objects so
that these experiments are essentially limited by the accuracy in the
knowledge of the Casimir force. References on these topics as well as
further discussions of the motivations for studying the Casimir force may be
found in \cite{Reynaud01,Genet03}.

Casimir considered an ideal configuration with two perfectly reflecting
mirrors in vacuum. But the experiments are performed with real reflectors,
for example metallic mirrors which show perfect reflection only at
frequencies below the plasma frequency characterizing the metal.
Accounting for imperfect
reflection and its frequency dependence is thus essential for obtaining a
reliable theoretical expectation of the Casimir force in a real situation.
This is also true for other corrections to the ideal Casimir formula
associated with the experimental configuration: experiments are
performed at room temperature, with the effect of thermal fluctuations
superimposed to that of vacuum fluctuations; in most experiments, the force
is measured between a plane and a sphere, and not between two
parallel planes; also, the surface state of the plates, in particular their
roughness, should affect the force.
Here, we will focus our attention on the evaluation of the
Casimir force between two plane mirrors with arbitrary frequency dependent
reflection amplitudes and ignore the other corrections associated with
the effect of a non zero temperature or a non plane geometry.

The consideration of real mirrors is important not only for the analysis of
experiments but also from a conceptual point of view. Real mirrors are
certainly transparent at the limit of high frequencies and this allows one
to dispose of the divergences associated with the infiniteness of vacuum
energy. This point was already alluded to in Casimir's papers and an
important step in this direction was the Lifshitz theory of the Casimir
force between two dielectric bulks \cite{Lifshitz56,Schwinger78}.
Here we will use the general expression of the Casimir force obtained
for two plane mirrors characterized by arbitrary frequency dependent
reflection amplitudes \cite{Jaekel91}.
This expression is directly associated with an interpretation of the force
as resulting from the radiation pressure exerted by vacuum fluctuations upon
the two mirrors which form a Fabry-Perot cavity.
The balance between repulsive and attractive contributions associated with
resonant and antiresonant frequencies gives the net Casimir force.
This method always leads to a finite result as a consequence of the causality
properties and high-frequency transparency obeyed by any real mirror.
In other words, the properties of real mirrors are sufficient to obtain
a regular expression of the Casimir force, in spite of
the infiniteness of vacuum energy.

In the next section, we will first recall this general expression of the
Casimir force between real mirrors and briefly discuss its application to
theory-experiment comparison. We will then focus our attention on the simplest
model of metallic mirrors, namely bulk mirrors with the optical response of
metals described by the plasma model. This model is not sufficient for an
accurate evaluation of the Casimir force but its simplicity allows one
to discuss qualitatively a lot of interesting physical features.
In particular, we will discuss the analogies and differences between the
results obtained with this model for the Casimir force between macroscopic
mirrors
and for the Casimir-Polder force between two atoms in vacuum. In both cases,
simple power laws are obtained at the limits of long and short distances.
The indices of these power laws are different in the two cases but the
crossovers between short and long-distance laws present profound analogies.
Let us briefly mention the convention we usually use to name the different
interaction
ranges (there is no overall accepted denomination yet). While we name
the long range interaction between atoms usually Casimir-Polder force, as
Casimir
and Polder were the first to derive the correct expression in the retarded
intercation limit,
we call the short distance limit of the interaction between atomic bodies
the van der Waals
or London interaction, after the two physicists the first of whom predicted
phenomenologically
while the second derived quantum mechanically the correct interaction law.

Furthermore, we will show that the Casimir force at short distances can
be understood as the London interaction between the elementary excitations of
both scatterers, that is the surface plasmons of
the two bulk mirrors. These surface plasmons are oscillating
electromagnetic fields, strongly
localized at the surface of a metal (evanescent waves) and
associated with the collective motion of electrons.

Before entering this discussion, we may emphasize that the existence
of vacuum fluctuations alters the physical conception of empty space~:
in contrast to classical physics, quantum theory forces us to consider
vacuum as permanently filled by field fluctuations having observable effects.
Van der Waals and Casimir forces are nothing but the effects of vacuum
radiation pressure on microscopic or macroscopic objects at rest.
Vacuum radiation pressure also induces dynamical effects for objects
moving in vacuum and these effects are directly connected to the problem
of relativity of motion. Discussions of these dynamical effects and
references may be found in \cite{Lambrecht03}.

\section{Casimir force between real mirrors}

We now come to a more precise discussion of the Casimir force.
Casimir calculated this force in a geometrical configuration where two plane
mirrors are placed a distance $L$ apart from each other, parallel to each
other, the area $A$ of the mirrors being much larger than the squared
distance $\left( A\gg L^{2}\right) $. Considering the ideal case of perfectly
reflecting mirrors Casimir obtained the following expression for the force
\begin{equation}
F_{\rm Cas}= \frac{\hbar c\pi ^2 A}{240 L^4} \label{FCasimir}
\end{equation}
We have chosen the sign convention found in most papers on Casimir force
with a positive value of $F_{\rm Cas}$ corresponding to an attraction,
that is also a negative pressure.

This ideal Casimir formula only depends on geometrical quantities $A$ and
$L$ and
on two fundamental constants, the speed of light $c$ and Planck constant
$\hbar$.
This remarkable universal feature corresponds to the fact that the optical
response of perfect mirrors is saturated~: mirrors cannot reflect more than
100 \% of the incoming light, whatever their atomic constitution may be.
This makes an important difference between ideal Casimir forces and the Van
der
Waals forces, discussed below, which depend on atomic polarizabilities.
Now experiments are performed with metallic mirrors which do not reflect
all field frequencies perfectly. This has certainly to be taken into account
in the comparison between theoretical estimations and experimental
measurements.
This also entails that the Casimir force between real mirrors depends on
the atomic
structure constants which determine the optical properties of the latter.

Imperfectly reflecting mirrors will be described by scattering amplitudes which
depend on the frequency, wavevector and polarization while obeying general
properties of stability, high-frequency transparency and causality. The two
mirrors form a Fabry-Perot cavity with the consequences well-known in
classical or quantum optics~: the energy density of the intracavity field is
increased for the resonant frequency components whereas it is decreased for
the non resonant ones. The Casimir force is but the result of the balance
between the radiation pressure of the resonant and non resonant modes which
push the mirrors respectively towards the outer and inner sides of the
cavity \cite{Jaekel91}. This balance includes not only the contributions of
ordinary waves propagating freely outside the cavity but also that of
evanescent waves. These two sectors of ordinary and evanescent waves are
directly connected by analyticity properties of the scattering amplitudes
(see a more precise discussion below).

The Casimir force may then be written as an integral over frequencies $\omega$,
transverse wavevectors ${\bf k}$ and a sum over polarizations $p$ of the
vacuum field modes. Due to the analycity properties, the integral may be
written also over imaginary frequencies $\omega = i \xi$ (with $\xi$ real)
\begin{eqnarray}
&&F =\hbar A\sum_{p}\int \frac{{\rm d}^{2}{\bf k}}{4\pi ^{2}}%
\int_{0}^{\infty }\frac{{\rm d}\xi }{2\pi }\ 2\kappa \ \frac{\rho _{{\bf k}}
^{p}\left[ i\xi \right] }{1-\rho _{{\bf k}}^{p}\left[ i\xi \right] }
\label{Force} \\
\rho _{{\bf k}}^{p}\left[ i\xi \right] &=& r_{{\bf k},1}^{p}\left[ i\xi
\right]
r_{{\bf k},2}^{p}\left[ i\xi \right] e^{-2\kappa L}  \qquad , \qquad
\kappa =\sqrt{{\bf k}^{2}+ \frac{\xi ^{2}}{c^2}}  \nonumber
\end{eqnarray}
In this expression, $\rho$ represents the multiplication factor for the field
after a round trip in the cavity~: it is the product of the reflection
amplitudes
$r_{1}$ and $r_{2}$ of the two mirrors and of an exponential phase factor.
The fraction $\frac{\rho}{1-\rho}$ is the sum of similar factors
over the number of round trips inside the Fabry-Perot cavity
\begin{eqnarray}
\frac{\rho}{1-\rho} &=& \rho + \rho ^2 + \rho ^3 + \ldots
\end{eqnarray}
In other words, $\rho$ is the `open loop function' associated with the cavity
while $\frac{\rho}{1-\rho}$ is the `closed loop function' taking into account
the feedback; all these quantities are evaluated at imaginary frequencies
through an analytical prolongation from their values at real frequencies.

Expression (\ref{Force}) holds for dissipative mirrors as well as for non
dissipative
ones \cite{Genet02}. It is regular for any frequency dependence of the
reflection
amplitudes obeying natural physical conditions~: causality of the amplitudes
and high-frequency transparency for each mirror, stability of the closed loop
function associated with the Fabry-Perot cavity.
Expression (\ref{Force}) tends towards the ideal Casimir formula
(\ref{FCasimir})
as soon as the mirrors are nearly perfect for the modes contributing to the
integral.
Incidentally, these results show that the Casimir force between two plane
mirrors
can be evaluated without any renormalization technique~: as guessed
by Casimir in his original paper, the properties of real mirrors themselves are
sufficient to enforce regularity of the Casimir force.

The condition under which the ideal Casimir result is approached will be
specified
below for bulk mirrors described by the plasma model.
More generally, the reduction of the Casimir force (\ref{Force}) with
respect to
the ideal formula (\ref{FCasimir}) due to the imperfect reflection of mirrors
is described by a factor
\begin{equation}
\eta _{\rm F} = \frac{F}{F_{\rm Cas}}  \label{etaF}
\end{equation}

We may proceed analogously for discussing the Casimir energy
between real mirrors, evaluated by integrating the force
\begin{equation}
E=-\int_{L}^{\infty }\ F\left( L' \right) \ {\rm d}L'
\end{equation}
For the energy, we use the standard convention of thermodynamics,
with a negative value corresponding to a binding energy \footnote{
This convention is opposite to that used in our papers quoted in the list
of references but it is better adapted to the forthcoming discussions.}.
The force (\ref{FCasimir}) between perfect mirrors is thus translated to
\begin{equation}
E_{\rm Cas}= -\frac{\hbar c\pi ^{2}A}{720L^{3}}  \label{ECasimir}
\end{equation}
Meanwhile, the force (\ref{Force}) between real mirrors corresponds to
the energy
\begin{equation}
E=A\sum_{p}\int \frac{{\rm d}^{2}{\bf k}}{4\pi ^{2}}\int_{0}^{\infty }
\frac{{\rm d}\xi }{2\pi }\ \hbar \ \log \left( 1-\rho _{{\bf k}}^{p}
\left[ i\xi \right] \right)      \label{Energy}
\end{equation}
This energy has its absolute value reduced by the effect of imperfect
reflection and the reduction is conveniently described by a factor
\begin{equation}
\eta _{\rm E} = \frac{E}{E_{\rm Cas}}  \label{etaE}
\end{equation}
This factor plays an important role in the discussion of the most
precise recent experiments.

In these experiments, the force has been measured between a sphere
and a plane and not between two plane parallel mirrors.
The Casimir force in this geometry is estimated from the proximity
force approximation \cite{Deriagin68} which amounts to integrate the
contributions of the various inter-plate distances as if they were
independent.
Although the accuracy of this approximation remains to be mastered,
it is usually thought to give a reliable approximation.
The force in the sphere-plane geometry is then given by the radius $R$
of the sphere and by the Casimir energy as evaluated in the plane-plane
configuration for the distance $L$ of closest approach
\begin{eqnarray}
F_{\rm sphere-plane} &=& \frac {2\pi R}{A}  \left| E_{\rm plane-plane} \right|
\end{eqnarray}
Using the results (\ref{ECasimir}-\ref{etaE}) obtained in the plane-plane
geometry, one finally obtains the Casimir force in the sphere-plane geometry
\begin{eqnarray}
F_{\rm sphere-plane} &=& \frac{\hbar c \pi^3 R}{360 L^3}  \eta _{\rm E}
\label{pft}
\end{eqnarray}
This theoretical expression is used for comparison with recent precise
measurements of the force \cite{Bordag01}.
It accounts for the effects of imperfect reflection and plane-sphere geometry
which have a significant impact on the value of the Casimir force.
This is not the case for the effects of thermal fluctuations and surface
roughness which only have a marginal influence ($<1\%$) in the same
experiments.
As a consequence, the good agreement ($\simeq 1\%$) obtained in the
theory-experiment comparison can be considered as a confirmation of the
existence and properties of the Casimir force as well as a test of the
corrections associated  with imperfect reflection and plane-sphere geometry.

We want to emphasize that a precise description of the optical response of the
mirrors is necessary to reach an accurate evaluation of the effect of
imperfect reflection. If an accuracy of the order of 1\% is aimed at,
this description should take into account the knowledge of optical
data of the metals on a wide frequency range \cite{Lambrecht00}.
In particular, the plasma model is not sufficient for such an accurate
evaluation. As announced in the Introduction however, we will use this
simple model in the following to get interesting results about the
comparison of Casimir and Casimir-Polder forces.

\section{Reflection on bulk mirrors described by the plasma model}

For bulk mirrors, the reflection amplitudes are simply given by the Fresnel
laws corresponding to the vacuum/metal interface with different expressions
for the two polarizations TE and TM
\begin{eqnarray}
&&r^{\rm TE} = \frac{\kappa-\kappa_{\rm m}}{\kappa+\kappa_{\rm m}}
\qquad ,\qquad
r^{\rm TM} = \frac{\kappa_{\rm m}-\varepsilon_{\rm m}\kappa}
  {\kappa_{\rm m}+\varepsilon_{\rm m}\kappa} \nonumber \\
&&\kappa_{\rm m} =\sqrt{{\bf k}^{2}+ \varepsilon_{\rm m} \frac{\xi
^{2}}{c^2}}
\qquad ,\qquad
\kappa=\sqrt{{\bf k}^{2}+\frac{\xi ^{2}}{c^2}}  \label{reflAmpl}
\end{eqnarray}
$\varepsilon_{\rm m}$ is the dielectric function of the metal
given by the plasma model
\begin{equation}
\varepsilon _{\rm m} \left[ i\xi \right] = 1 + \frac{\omega_{\rm P}^2}{\xi ^2}
\qquad ,\qquad \lambda _{\rm P}=\frac{2\pi c}{\omega _{\rm P}}
\end{equation}
with $\omega_{\rm P}$ the plasma frequency and $\lambda_{\rm P}$ the plasma
wavelength;
$\kappa$ is the quantity already defined in (\ref{Force}); $\kappa_{\rm m}$ is
the expression defined analogously in the metal with the dielectric function
$\varepsilon_{\rm m}$.

These expressions are well known but their analyticity properties deserve
a special attention in the context of the present discussion.
Causality entails that the reflection amplitudes are analytical functions
of the frequency in the `physical domain' defined in the complex plane
by a positive real part for $\xi$ \cite{Landau}
\begin{equation}
\omega \equiv i\xi \qquad ,\qquad \Re\xi >0    \label{physicalDomain}
\end{equation}
Analyticity has to be understood with ${\bf k}$ and $p$ fixed,
the branch of the square roots being taken so that
\begin{equation}
\Re\kappa_{\rm m} >0 \qquad ,\qquad \Re\kappa>0  \label{kappaBranch}
\end{equation}
The sectors of ordinary and evanescent waves lie on the boundary
of this domain~: they indeed correspond to real frequencies $\omega$,
that is also to purely imaginary values for $\xi$.
They are distinguished by the values of $\kappa$ which are
purely imaginary for ordinary waves ($\omega \geq c\left| {\bf k}\right|$),
but real for evanescent waves ($\omega < c\left| {\bf k}\right|$).
In the latter case, $\kappa$ is just the inverse of the penetration
length in vacuum of waves coming from the refracting medium.

As already alluded to, analyticity also connects these two sectors to the
sector
of imaginary frequencies ($\xi$ real). Precisely, expressions of the
reflection
amplitudes or of the loop functions in this sector are obtained from similar
expressions written for real frequencies through an analytical continuation.
This property was in fact used to write the force (\ref{Force}) and energy
(\ref{Energy}) as integrals over imaginary frequencies \cite{Genet02}.
Note in particular that the exponential factor $\exp\left(-2\kappa L\right)$
appearing in the loop function $\rho$ describes the frustration of total
reflection
on each vacuum/metal interface due to the evanescent propagation of the field
through the length $L$ of the cavity. This explains why the radiation pressure
of vacuum modes is not identical on the internal and external sides of each
mirror and, therefore, why the Casimir force has finally a non null value.

It is also worth discussing in more details the modulus of the reflection
amplitude which is expected to be smaller than unity as a consequence of
unitarity of the scattering on a mirror
\begin{equation}
\left| r \right| \leq 1   \label{Passiv}
\end{equation}
This property is certainly true in the sector of ordinary waves where it
is effectively a direct consequence of unitarity, for lossy as well as
lossless mirrors \cite{Genet02}.
It is easily seen (\ref{reflAmpl}) that it is also true in the sector of
imaginary
frequencies where $\kappa$, $\kappa_{\rm m}$ and $\varepsilon_{\rm m}$ are real
and positive. But the case of evanescent waves requires a closer examination
and the result of this examination depends on the polarization.

For TE waves, it follows from (\ref{kappaBranch}) that (\ref{Passiv}) holds
also in the evanescent sector. It is then a consequence of the
Phragm\'en-Lindel\"of theorem \cite{Phragmen} that (\ref{Passiv}) is true
on the whole physical domain (\ref{physicalDomain}). Then the open loop
function $\rho$ also has its modulus smaller than unity in this domain
which ensures stability of the closed loop gain $\frac{\rho}{1-\rho}$.
This stability means that neither the mirrors forming the cavity nor the
vacuum fields enclosed in this cavity have the ability to sustain the
oscillation which would be associated with a pole of $\frac{\rho}{1-\rho}$
in the physical domain \cite{Lambrecht97}.
For TM waves, this stability property is also true and no self-sustained
oscillation of the Fabry-Perot cavity occurs. However, (\ref{Passiv})
does not hold in the evanescent sector since the reflection amplitudes
reach large values corresponding to the existence of surface plasmon
resonances (see a more precise discussion below). This means that
the closed loop amplitude is still stable although the open loop
amplitude has a modulus larger than unity. In more mathematical terms,
the oscillation condition $\rho=1$, which corresponds to a pole of
the closed loop function $\frac{\rho}{1-\rho}$, is never met in the
physical domain (\ref{physicalDomain}) although $\left| \rho \right|$
may exceed unity in this domain.

This discussion can be made more precise for bulk mirrors
described by the plasma model. To this aim, we first consider reflection
of TM waves on one such mirror. The corresponding reflection amplitude
$r^{\rm TM}$ given by (\ref{reflAmpl}) is seen to diverge when
$\kappa _{\rm m}+\varepsilon _{\rm m}\kappa =0$, which
defines the surface plasmon resonance condition \cite{Barton79}.
With the plasma model, this condition can be solved as an
expression for the frequency in terms of the transverse wavevector
\footnote{
Note that the condition $\kappa _{\rm m}-\varepsilon _{\rm m}\kappa=0$,
differing by a mere sign, corresponds to a vanishing reflection amplitude
and defines the Brewster frequency with the following expression
$\omega_{\rm Brewster} ^{2}=\frac{\omega _{\rm P}^{2}+2c^{2}{\bf k}^{2} +
\sqrt{\omega_{\rm P}^{4}+4c^{4}{\bf k}^{4}}}{2}$.}
\begin{equation}
\omega_{\rm plasmon} ^{2}=\frac{\omega _{\rm P}^{2}+2c^{2}{\bf k}^{2}-
\sqrt{\omega_{\rm P}^{4}+4c^{4}{\bf k}^{4}}}{2}
\label{freqPlasmon}
\end{equation}
The surface plasmon frequency $\omega_{\rm plasmon}$ is real and lies on the
boundary $\Re\xi=0$ of the physical domain (\ref{physicalDomain}).
When dissipation is taken into account, for example by considering the
Drude model instead of the plasma model, the pole of the closed loop
amplitude is pushed from this boundary $\Re\xi=0$ into the unphysical
domain $\Re\xi<0$. It follows that the divergence encountered with the
plasma model is transformed into a resonance. In the vicinity of this
resonance, the modulus $\left| r^{\rm TM} \right|$ of the reflection
amplitude exceeds unity but the closed loop function
remains stable since its pole lies outside the physical domain.

When two mirrors are considered, their surface plasmons are coupled by
evanescent propagation through the cavity so that their frequencies
are displaced. Their displacement can be seen as responsible for the
interaction between the two mirrors \cite{VanKampen68,Schram73}.
This approach will be discussed in more detail below since it leads to an
interesting interpretation of the Casimir force at short distances.

\section{Power laws in the limits of long and short distances}

\begin{figure}[tbh]
 \centering
\includegraphics[width=6cm]{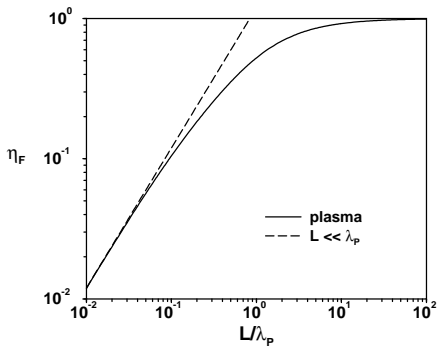}
\caption{The solid curve represents the variation of the force reduction factor 
$\eta _{\rm F}$ as a function of the reduced distance $\frac{L}{\lambda_{\rm P}}$~:
$\eta _{\rm F}$ goes to unity at large distances whereas it varies linearly
with $\frac{L}{\lambda_{\rm P}}$ at short distances; the latter behaviour is
represented by the dashed line.}
\label{FigPlasmaModel}
\end{figure}

Using the simple mirror model of the preceding section,
we now discuss the relation between the effect of imperfect reflection
and the explored distance range.
The evaluation of the reduction factor (\ref{etaF}) for bulk mirrors
described by the plasma model has been presented in \cite{Lambrecht00,Genet00}.
In this simple case, the reduced force $\eta _{\rm F}$ is a function of a
single parameter, namely the reduced distance $\frac{L}{\lambda_{\rm P}}$.
We draw this function on Figure \ref{FigPlasmaModel}.

We first note that the factor $\eta _{\rm F}$ gets close to unity
at large distances
\begin{eqnarray}
&&L \gg \lambda _{\rm P} \qquad , \qquad \eta _{\rm F}\simeq 1 \nonumber \\
&&F\simeq \frac{\hbar cA\pi ^{2}}{240L^{4}}
\label{longDist}
\end{eqnarray}
This corresponds to a well known interpretation~: large distances $L \gg
\lambda _{\rm P}$
correspond to low frequencies $\omega \ll \omega _{\rm P}$ for which
metallic mirrors
described by the plasma model are nearly perfect reflectors;
this is why the ideal Casimir formula (\ref{FCasimir}) is a very good
approximation of the real force (\ref{Force}) in this limit.

Otherwise, the factor $\eta _{\rm F}$ is smaller than unity
and describes the reduction of the force due to the imperfect
reflection of the metallic mirror at high frequencies.
In the limiting case of distances small with respect to the plasma
wavelength, the reduction becomes quite significant since the factor
$\eta _{\rm F}$ varies linearly with the small factor $\frac{L}{\lambda
_{\rm P}}$
\begin{eqnarray}
&&L\ll \lambda _{\rm P}\qquad , \qquad \eta _{\rm F}\simeq \alpha
\frac{L}{\lambda _{\rm P}} \nonumber \\
&&F\simeq \alpha
\frac{\hbar cA\pi ^{2}}{240\lambda _{\rm P}L^{3}}
\label{shortDist}
\end{eqnarray}
The dimensionless constant $\alpha$ was evaluated numerically
in \cite{Lambrecht00,Genet00} and it will be given a more detailed
interpretation in the following.

At this point, it is worth comparing the variation with distance of the
Casimir force with that of the Van der Waals force between two atoms in vacuum.
Casimir and Polder \cite{CasimirP48} indeed showed that the latter force obeys
power laws in the two limits of short and long distances, with the exponent
being
changed by one unit when going from one limit to the other and the crossover
taking place when the interatomic distance $L$ crosses the typical atomic
wavelength $\lambda _{\rm A}$.
The same behaviours are also observed for the Casimir force between two
metallic
mirrors with the plasma wavelength $\lambda _{\rm P}$ playing the role of
$\lambda _{\rm A}$.

Since this comparison allows one to get interesting insight on the Casimir
force,
we briefly recall the results known for the Casimir-Polder force in the next
section and then discuss the analogies and differences between both cases.

\section{Reminders~: the Casimir-Polder force between two atoms}

In the present section we remind a few interesting results about the Van
der Waals
force between two atoms. We first rewrite the general expression obtained by
Casimir and Polder \cite{CasimirP48} as an interaction energy \cite{PowerT93}
\begin{eqnarray}
E_{\rm CP}&=&-\frac{\hbar c}{\pi L^{2}}\int_{0}^{\infty }{\rm d}\kappa \
\alpha _{\rm A}^{2}\left[ ic\kappa \right] \nonumber \\
&\times& \left( \kappa ^{4}+\frac{%
2\kappa ^{3}}{L}+\frac{5\kappa ^{2}}{L^{2}}+\frac{6\kappa }{L^{3}}+\frac{3}{%
L^{4}}\right) e^{-2\kappa L}  \nonumber \\
&&\alpha _{\rm A}\left[ ic\kappa \right] =\sum_{n>0}\frac{E_{n}A_{n}}{%
E_{n}^{2}+\hbar ^{2}c^{2}\kappa ^{2}}
\label{CasimirPolder}
\end{eqnarray}
The interaction energy is an integral over imaginary frequencies
$\omega=ic\kappa$,
$L$ is the distance between the two atoms and $\alpha _{\rm A}$
represents the frequency dependent polarizability of the atoms \footnote{
Expression (\ref{CasimirPolder}) is written for two identical atoms;
otherwise, $\alpha _{\rm A}^{2}$ should be replaced by the product of the
polarizabilities of the two atoms.}. The polarizability is written here in
the dissipation free approximation and it depends on the energy $E_{n}$ of
the $n-$th atomic state measured with respect to the ground state and on a
coefficient $A_{n}$ proportional to the square modulus of the dipole matrix
element associated with the transition. In the sequel of the paper,
we discuss the profound analogies as well as a few significant differences
appearing in the comparison of expression (\ref{CasimirPolder}) and
(\ref{Energy}).

We want first to stress a fundamental analogy related to the very significance
of long-range interactions such as Casimir-Polder or Casimir effects.
Feinberg and Sucher \cite{FeinbergS70} have written a still more general
expression of the Casimir-Polder energy with magnetic polarizabilities taken
into account besides the electric polarizabilities which appear in
(\ref{CasimirPolder}). Since magnetic terms are smaller than electric ones,
we ignore them here but however refer to their paper for a very general
discussion of long-range interactions which decrease as power laws of the
distance. This makes an important difference with molecular interactions
which decrease as exponential laws and, for this reason, dominate at typical
atomic distances. An interesting fact is that long range interactions, in
contrast to molecular interactions, may be deduced from the scattering
amplitudes evaluated for the two scatterers separately. In more formal words,
the two-centers scattering amplitudes may be written in terms of the
one-center
scattering amplitudes evaluated on shell for the photons and of free field
propagation factors. For example, the Casimir-Polder energy
(\ref{CasimirPolder}) between atoms is written in terms of the
polarizabilities
$\alpha _{\rm A}$ while the Casimir energy (\ref{Energy}) between mirrors
is written in terms of the reflection amplitudes $r$. In both cases,
free field propagation between the two centers is described by the exponential
factor $\exp\left(2\kappa L\right)$ where $2\kappa$ is the momentum
transfered on each scattering. In both cases too, the interaction energy
can be written equivalently as an integral over real or imaginary
frequencies, the equivalence being a consequence of causality properties.

These fundamental analogies should be appreciated in contrast with some
significant
differences. In particular, the atoms are point-like scatterers well
coupled to the
spherical waves centered on them whereas mirrors are plane scatterers which
fit the
definition of plane waves. Hence, the expression (\ref{Energy}) of the
Casimir force
shows explicitly the summation over transverse wavevectors ${\bf k}$ and
polarizations
$p$ while the field propagation directions and polarizations have been
traced over
in expression (\ref{CasimirPolder}) of the Casimir-Polder energy.
It also follows from the point-like character of atoms that their mutual
coupling
through the field is less efficient than for mirrors.
In other words, the two atoms form a poor-finesse cavity so that the higher
order
interferences terms, which play an important role in the Fabry-Perot cavity,
can be disregarded in the two-atoms problem.
This difference is made explicit by expanding the logarithm in
(\ref{Energy}) as
\begin{eqnarray}
\log \left( 1- \rho \right) &=& - \left( \rho + \frac{\rho ^2}{2}
+ \frac{\rho ^3}{3} + \ldots \right) \label{logExpand}
\end{eqnarray}
The lowest-order term in this expansion varies as $e^{-2\kappa L}$ with
distance
and thus corresponds to the term present in (\ref{CasimirPolder}). In
contrast,
the higher-order terms appearing in (\ref{logExpand}) and, therefore, in
(\ref{Energy})
are not accounted for in the perturbative expression (\ref{CasimirPolder}).

We now go one step further in the discussion of (\ref{CasimirPolder}) by
considering the large and short distance limits.
In the large distance limit where $L$ is greater than the wavelengths of the
various atomic transitions, retardation effect plays a dominant role. This
means that the exponential factor $e^{-2\kappa L}$ restricts significant
contributions
to the integral (\ref{CasimirPolder}) to low values of $\kappa$ for which the
polarizability remains nearly equal to its static value
$\alpha _{\rm A}\left[ 0 \right]$.
Hence, the interaction energy is obtained by evaluating a universal integral
\begin{eqnarray}
&&E_{\rm CP} =
- \frac{23}{4}\frac{\hbar c}{\pi L^{7}}\alpha _{\rm A}^{2}\left[ 0\right]
\qquad ,\qquad L \gg \lambda _{\rm A}  \label{VdWCP}\\
&&\frac{23}{4} \equiv
\int_{0}^{\infty }{\rm d}u\ \left( u^{4}+2u^{3}+5u^{2}+6u+3\right) e^{-2u}
\nonumber 
\end{eqnarray}
This result bears some similarity with the Casimir energy evaluated at large
distances which depends only on the static value $r\left[ 0 \right]$
of reflection amplitudes. In the Casimir case, this value goes to unity
and then completely disappears from the ideal Casimir formula
(\ref{ECasimir}).
In the Casimir-Polder case in contrast, $\alpha _{\rm A}\left[ 0 \right]$
determines the global magnitude of the interaction energy. The difference
between the two power laws can be attributed to dimensional arguments
with $\alpha _{\rm A}$ having the dimension of a volume.

We then consider the short distance limit $L\ll \lambda _{\rm A}$ where the
retardation effect is negligible. This means that the exponential factor may
be discarded and also entails that the last term in the parenthesis appearing
in (\ref{CasimirPolder}) dominates the other ones. The interaction energy then
scales as $\frac{1}{L^6}$, a result well known for the Van der Waals
interaction
calculated by London with retardation effects ignored \cite{London30}
\begin{equation}
E_{\rm CP}=-\frac{3\hbar c}{\pi L^{6}}\int_{0}^{\infty }{\rm d}\kappa \
\alpha _{\rm A}^{2}\left[ ic\kappa \right] \qquad ,\qquad
L\ll \lambda _{\rm A}
\label{VdWLondon}
\end{equation}
Using the expression of the frequency-dependent polarizability
(see \ref{CasimirPolder}) and the integral
\begin{eqnarray}
&&\int_{0}^{\infty }\frac{a}{a^{2}+x^{2}}\frac{b}{b^{2}+x^{2}}{\rm d}x=\frac{%
\pi }{2}\frac{1}{a+b} \nonumber \\
&& a,b>0  \label{integralVdW}
\end{eqnarray}
one rewrites the London expression as
\begin{equation}
E_{\rm CP}=-\frac{3}{2L^6}
\sum_{n,n^\prime > 0}\frac{A_{n}A_{n^\prime}}{E_{n}+E_{n^\prime}}
\qquad ,\qquad L \ll \lambda _{\rm A}
\end{equation}

At this point we may stress that the change of exponent in the power laws
is effectively similar in the Casimir and Casimir-Polder cases~: the
Casimir energy
scales as $\frac{1}{L^3}$ at large distances and $\frac{1}{L^2 \lambda_{\rm
P}}$
at short distances while the Casimir-Polder energy scales as
$\frac{1}{L^7}$ at large
distances and $\frac{1}{L^6 \lambda_{\rm A}}$ at short distances. Keeping
in mind
that the global change of exponents is explained by dimensional arguments,
the change
of exponent at the crossover is effectively the same.
In order to stress this point, we could introduce a $\eta$
factor in the Casimir-Polder case as the ratio of the general expression
(\ref{CasimirPolder}) to the long-distance expression (\ref{VdWCP})
\begin{eqnarray}
\eta ^{\rm CP} _{\rm E} &=& \frac{4}{23} \int_{0}^{\infty }{\rm d}K \
\frac{\alpha _{\rm A}^{2}\left[ ic\kappa \right]}{\alpha _{\rm A}^{2}\left[
0 \right]} \nonumber \\
&\times&\left( K^4 + 2K^3 + 5K^2 + 6K + 3 \right) e^{-2K}  \nonumber \\
K &\equiv& \kappa L
\end{eqnarray}
This factor goes to unity at large distances and varies linearly with
$L\ll \lambda _{\rm A}$ at short distances; it thus varies roughly as
the $\eta$ factor defined above for the Casimir problem.

In the next section, we show that this analogy may be pushed one step
further, allowing one to give an interesting interpretation of the Casimir
force
at short distances as resulting from the London (non retarded) interaction
between
the elementary excitations in the two scatterers, that is the surface plasmons
which live at the interface between each bulk mirror and vacuum
\cite{Barton02}.

\section{London interaction between surface plasmons}

In this last section, we study the Casimir energy between metallic plates
described
by the plasma model at the limit of short distances $L\ll \lambda _{\rm P}$.

To this aim, we first show that its expression (\ref{Energy}) can be greatly
simplified at this limit. Indeed, values of $\kappa$ contributing
significantly to
the integral correspond to $\kappa L$ of the order of unity~: large values have
their contribution suppressed by the exponential factor $\exp\left(-2\kappa
L\right)$
while small values correspond to a small measure in the integration over
transverse
wavevector. Since $L\ll \lambda _{\rm P}$, this condition also means that
$\kappa \lambda _{\rm P} \gg 1$. Meanwhile, a non vanishing value of $r$
implies $\xi \lesssim \omega_{\rm P}$, that is also
$\frac{\xi }{c}\lambda _{\rm P} \lesssim 1$.
In these conditions, $\kappa$ and $\kappa _{\rm m}$ are both approximately
equal to $\left| {\bf k} \right|$, the TE reflection amplitude is negligible
whereas the TM reflection amplitude takes a simple Lorentzian form
\begin{equation}
r^{\rm TM} \simeq \frac{\omega_{\rm S}^{2}}{\omega ^{2}-\omega _{\rm S}^{2}}
\qquad ,\qquad \omega _{\rm S}^2 \equiv \frac{\omega _{\rm P}^{2}}{2}
\end{equation}
The frequency $\omega _{\rm S}$ is the constant value of the surface plasmon
frequency $\omega_{\rm plasmon}$ obtained by putting the condition
$\left| {\bf k} \right| \lambda _{\rm P} \gg 1$ in (\ref{freqPlasmon}).

For a cavity made with two identical mirrors, the open loop function then takes
the simple form
\begin{equation}
\rho ^{\rm TM} = \left( \frac{\omega _{\rm S}^2}{\omega ^2-\omega _{\rm
S}^2} \right)
^{2} e^{-2 \left| {\bf k} \right| L}
\label{rhoTM}
\end{equation}
The poles of the closed loop function, given by the condition $\rho ^{\rm
TM} =1$,
therefore correspond to the frequencies
\begin{equation}
\omega _{{\rm S} \pm} =\omega _{\rm S}\sqrt{1 \pm e^{- \left| {\bf k}
\right| L}}
\end{equation}
This expression shows how the surface plasmons corresponding to the two
mirrors are
displaced from their original frequencies $\omega _{\rm S}$ due to their
coupling
through the cavity.
It also provides an interesting interpretation of the Casimir interaction
energy
which can be written from (\ref{rhoTM}) as
\begin{eqnarray}
E &=& A \int \frac{{\rm d}^{2}{\bf k}}{4\pi ^2}\ \left(
\frac{\hbar \omega _{{\rm S} +}}{2} + \frac{\hbar \omega _{{\rm S} -} }{2}
-2\frac{\hbar \omega _{\rm S}}{2}\right)
\end{eqnarray}
The Casimir energy is nothing but the result of the shift of zero-point
energies
of the surface plasmons due to their coupling through the cavity field
\cite{Barton02}. This result has been obtained with retardation effects
neglected~:
this follows from the assumptions used in the derivation and is also
apparent in
the fact that the exponential factors depend on the transverse wavevectors but
not on the frequencies of the field.

We now conclude by giving an analytical form of this London-like expression of
the Casimir energy at short distances. We come back to equation (\ref{Energy})
that we expand with the help of (\ref{logExpand}). We ignore the vanishing TE
contribution and obtain
\begin{eqnarray}
&&E=- \hbar A \sum_{n=1}^{\infty }
\int \frac{{\rm d}^{2}{\bf k}}{4\pi ^{2}}
\frac {e^{-2n \left| {\bf k} \right| L}} n I_{n}
\nonumber \\
&&I_{n} = \int_{0}^{\infty} \frac{{\rm d}\xi }{2\pi}
\left( r^{\rm TM} \left[ i\xi \right] \right) ^{2n}
= \frac{\omega_{\rm S}}{4}
\frac{\left( 4n-3\right) !!}{\left( 4n-2\right) !!}
\nonumber \\
&&\frac{\left( 4n-3\right) !!}{\left( 4n-2\right) !!}
\equiv \frac{1.3.5...\left( 4n-3\right) }{2.4.6...\left( 4n-2\right) }
\end{eqnarray}
Collecting these results finally leads to the Casimir energy at short distances
\begin{equation}
E \simeq - \frac{\hbar cA}{16\sqrt{2} L^2 \lambda _{\rm P}}
\sum_{n=1}^{\infty }\frac{1}{n^{3}} \frac{\left( 4n-3\right) !!}{\left(
4n-2\right) !!}
\end{equation}

This result may be equivalently expressed in terms of the Casimir force or
of the force
reduction factor (\ref{shortDist}) with now a formal expression for the
numerical
coefficient $\alpha $
\begin{eqnarray}
\alpha &=& \frac{30}{\sqrt{2}\pi ^{2}}
\sum_{n=1}^{\infty }\frac{1}{n^{3}}
\frac{\left( 4n-3\right) !!}{\left( 4n-2\right) !!}
\nonumber\\
&=& \frac{15}{\sqrt{2}\pi ^{2}}
\left( 1 + \frac{5}{64} + \frac{7}{384} + \ldots \right)
\simeq 1.193
\end{eqnarray}
This reproduces the result of \cite{Lambrecht00,Genet00}.
Note that the integral $I_{1}$ corresponding to the lowest order term $n=1$
has the same form as the integral (\ref{integralVdW}) used in the
calculation of the
London expression (\ref{VdWLondon}), with the surface plasmon frequency
playing the role
of an atomic resonance frequency.
The other contributions $n>1$ represent the effect of higher-order
interferences in the
Fabry-Perot cavity. It is worth emphasizing that they have the same scaling
law versus
distance as the lowest order term $n=1$ and contribute significantly
($\sim 10\%$) to the final result.

\medskip
\noindent
Acknowledgements~: We thank G.~Barton and M.-T.~Jaekel for 
stimulating discussions.


\vskip 30pt
\begin{references}
\bibitem{Casimir48}  H.B.G. Casimir, {\it Proc. K. Ned. Akad. Wet.} {\bf 51}
(1948) 793.

\bibitem{Sparnaay89}  M.J. Sparnaay, in {\it Physics in the Making} eds
Sarlemijn A. and Sparnaay M.J. (North-Holland, 1989) 235
and references therein.

\bibitem{Milonni94}  P.W. Milonni, {\it The quantum vacuum} (Academic, 1994).

\bibitem{Mostepanenko97}  V.M. Mostepanenko and N.N. Trunov,
{\it The Casimir effect and its applications} (Clarendon, 1997).

\bibitem{LamoreauxResource99} S.K. Lamoreaux, Resource Letter in
{\it Am. J. Phys.} {\bf 67} (1999) 850.

\bibitem{Bordag01}  M. Bordag, U. Mohideen and V.M. Mostepanenko, {\it Phys.
Reports} {\bf 353} (2001) 1.

\bibitem{Lambrecht02}  A. Lambrecht and S. Reynaud, {\it S\'eminaire
Poincar\'e}
{\bf 1} (2002) 107.

\bibitem{Reynaud01}  S. Reynaud, A. Lambrecht, C. Genet and M.T. Jaekel,
{\it C. R. Acad. Sci. Paris} {\bf 2-IV} (2001) 1287 [arXiv:quant-ph/0105053].

\bibitem{Genet03}  C. Genet, A. Lambrecht and S. Reynaud,
in {\it On the nature of dark energy}, to appear (2003)
[arXiv:quant-ph/0210173].

\bibitem{Lifshitz56}  E.M. Lifshitz, {\it Sov. Phys. JETP} {\bf 2} (1956) 73.

\bibitem{Schwinger78}  J. Schwinger, L.L. de Raad Jr. and K.A. Milton,
{\it Ann. Physics} {\bf 115} (1978) 1.

\bibitem{Jaekel91}  M.T. Jaekel and S. Reynaud, {\it J. Physique} {\bf I-1}
(1991) 1395 [arXiv:quant-ph/0101067].

\bibitem{Lambrecht03}  A. Lambrecht, in the present volume (2003).

\bibitem{Genet02}  C. Genet, A. Lambrecht and S. Reynaud, preprint (2002)
[arXiv:quant-ph/0210174].

\bibitem{Deriagin68}  B.V. Deriagin, I.I. Abrikosova and E.M. Lifshitz,
{\it Quart. Rev.} {\bf 10} (1968) 295.

\bibitem{Lambrecht00}  A. Lambrecht and S. Reynaud, {\it Eur. Phys. J.}
{\bf D8} (2000) 309.

\bibitem{Landau} L.D. Landau and E.M. Lifshitz, {\it Elecrodynamics of
continuous media}
(Butterworth and Heinemann, 1998), chap. IX.

\bibitem{Phragmen}  See a discussion and references in \cite{Jaekel91}.

\bibitem{Lambrecht97}  A. Lambrecht, M.T. Jaekel and S. Reynaud, {\it Phys.
Lett.} {\bf A 225} (1997) 188 [arXiv:quant-ph/9801055].

\bibitem{Barton79}  G. Barton, {\it Rep. Prog. Phys.} {\bf 42} (1979) 65.

\bibitem{VanKampen68} N.G. Van Kampen, B.R.A. Nijboer and K. Schram,
{\it Phys. Lett.} {\bf 26A} (1968) 307.

\bibitem{Schram73} K. Schram, {\it Phys. Lett.} {\bf 43A} (1973) 282.

\bibitem{Genet00}  C. Genet, A. Lambrecht and S. Reynaud, {\it Phys.
Rev.} {\bf A62} (2000) 0121110.

\bibitem{CasimirP48}  H.B.G. Casimir and D. Polder, {\it Phys. Rev.}
{\bf 73} (1948) 360.

\bibitem{PowerT93}  E.A. Power and T. Thirunamachandran, {\it Phys. Rev.}
{\bf A48} (1993) 4761; {\bf A50} (1994) 3929.

\bibitem{FeinbergS70}  G. Feinberg and J. Sucher, {\it Phys. Rev.}
{\bf A2} (1970) 2395.

\bibitem{London30}  F. London, {\it Z. Phys.} {\bf 63} (1930) 245.

\bibitem{Barton02}  G. Barton, private communication (2002).

\end{references}
\end{document}